# Impact of Minority Carrier Lifetime on the Performance of Strained Ge Light Sources

David S. Sukhdeo[1], Krishna C. Saraswat[1], Birendra (Raj) Dutt[2,3], and Donguk Nam[*,4]

[1]*Department of Electrical Engineering, Stanford University, Stanford, CA 94305, USA*

[2]*APIC Corporation, Culver City, CA 90230, USA*

[3]*PhotonIC Corporation, Culver City, CA 90230, USA*

[4]*Department of Electronics Engineering, Inha University, Incheon 402-751, South Korea*

**\*Email: dwnam@inha.ac.kr**

**Abstract:** We theoretically investigate the impact of the defect-limited carrier lifetime on the performance of germanium (Ge) light sources, specifically LEDs and lasers. For Ge LEDs, we show that improving the material quality can offer even greater enhancements than techniques such as tensile strain, the leading approach for enhancing Ge light emission. Even for Ge that is so heavily strained that it becomes a direct bandgap semiconductor, the ~1 ns defect-limited carrier lifetime of typical epitaxial Ge limits the LED internal quantum efficiency to less than 10%. In contrast, if the epitaxial Ge carrier lifetime can be increased to its bulk value, internal quantum efficiencies exceeding 90% become possible. For Ge lasers, we show that the defect-limited lifetime becomes increasing important as tensile strain is introduced, and that this defect-limited lifetime must be improved if the full benefits of strain are to be realized. We conversely show that improving the material quality supersedes much of the utility of n-type doping for Ge lasers.

## *Introduction*

Germanium (Ge) light sources have garnered much attention for applications in both optical interconnects [1]–[5] and ultra-compact infrared sensing [6]–[9] due to Ge's inherent CMOS compatibility. Techniques such as tensile strain and n-type doping have been proposed to enhance the performance of Ge light emitters and extensive modeling exists for these approaches [10]–[13]. The importance of material quality, however, has been generally overlooked. This is a serious gap in the literature: epitaxial Ge typically suffers from high defect densities and poor carrier lifetimes [14]–[16], a problem which is well known to inhibit efficient light emission. Moreover, it is important for experimentalists to know whether or not this low carrier lifetime will present a performance bottleneck as Ge light sources mature and, if so, how severe this bottleneck will be. To address these questions, we will theoretically investigate the impact of the defect-limited carrier lifetime on the performance of strained Ge LEDs and lasers. For Ge LEDs, we will show that the achievable internal quantum efficiencies (IQE) are very low at present defect levels. Any improvement to the defect-limited lifetime will deliver a proportional improvement to the IQE, and the maximum possible IQE enhancement from reduced defect densities exceeds the predicted IQE enhancement from achieving a direct bandgap through tensile strain. Thus, we will show that the material quality is also an equally important obstacle to efficient Ge LEDs compared to Ge's indirect bandstructure. For Ge lasers, we show that the defect-limited lifetime is an increasingly problematic obstacle as tensile strain is introduced, and that this defect-limited lifetime must again be improved if the full benefits of strain are to be realized. More importantly, we will show that there is a synergy such that improving the material quality makes tensile strain an even more promising technique for improving Ge lasers. On the other hand, we will show that improving the material quality makes n-type doping a less useful technique for Ge lasers.

## *Band Structure Modeling & Carrier Statistics*

Our theoretical modeling process consists of various steps including: full bandstructure calculation, carrier statistics modeling, LED modeling, and laser modeling. The first step is to compute the full bandstructure of strained Ge over the intended range of strain values using $sp^3d^5s^*$ tight-binding following the approach of Refs [17], [18]. As illustrated in Fig. 1, the use of tight-binding allows us to compute not just the bandstructures' 2D cross sections (Fig. 1(a)) but also energies over a full 200x200x200 mesh of k-points encompassing the entirety of the first Brillouin Zone (Fig. 1(b)). This gives the full 4D bandstructure, i.e. energy as a function of the three wavevector components $k_x$, $k_y$ and $k_z$. We note that our tight-binding model assumes that Ge will become direct bandgap semiconductor at 2.4% biaxial tensile strain which means that our model is very conservative in that considerably more strain is needed to achieve a direct gap in our model than the ~1.7% predicted by most models [19]–[21]. Using our bandstructure model, we can also study general carrier statistics. To do this, we compute the occupancy probability given by Fermi-Dirac statistics for each allowed k-point and then integrate over the full k-point mesh of allowed energies encompassing the first Brillouin Zone to obtain the carrier concentration in each valley.

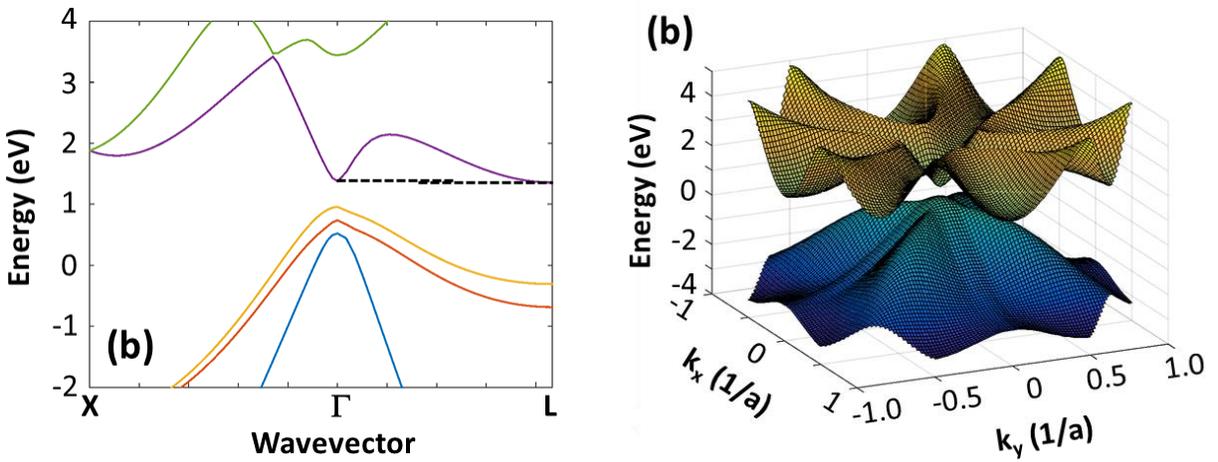

*Fig. 1.* *The bandstructure of Ge with 2.0% biaxial tensile strain computed by tight-binding. (a) 2D cross-sectional view. Black dashed horizontal lines are visual aids to help illustrate the Γ-L energy separation under tensile strain. (b) 3D cross-sectional view ($k_z$=0). The x- and y-components of the wavevector (**k**) are shown, with units given as multiples of the inverse lattice constant (1/a).*

## *LED Modeling*

Based on the bandstructure model and carrier statistics, it is possible to model the performance of a simple Ge LED. For this modeling, we will focus on the internal quantum efficiency (IQE) of a hypothetical Ge LED. Assuming a heterostructure design such that diffusion current is negligible and the injection efficiency will be virtually 100%, which can be achieved even in a simple Si/Ge/Si double heterostructure LED [22], the IQE will simply be the fraction of carrier recombination which is radiative. We can compute the radiative and non-radiative recombination rates by using Equation 1 and Equation 2, respectively. The radiative recombination rate of Equation 1 considers spontaneous emission from both the direct and indirect conduction valleys, with the radiative recombination lifetime being much faster in the direct valley, and so strain increases the radiative recombination rate by increasing the $\left(\frac{n_\Gamma}{n}\right)$ term. The non-radiative recombination rate of Equation 2, meanwhile, does not depend on whether electrons are in the direct or indirect conduction valley and is therefore independent of strain. Then, finally, the IQE is simply the ratio of radiative recombination to total recombination as given in Equation 3.

$$\begin{aligned} U_{\text{radiative}} &= R_L n_L p + R_\Gamma n_\Gamma p \\ &= R_L (n - n_\Gamma) p + R_\Gamma n_\Gamma p \\ &= R_L n p + (R_\Gamma - R_L) n_\Gamma p \end{aligned}$$

$$= R_L np + (R_\Gamma - R_L) np \left(\frac{n_\Gamma}{n}\right)$$

*Equation 1. Radiative recombination rate ($U_{radiative}$) in terms of the electron density (n), hole density (p) and fraction of electrons in the direct conduction valley $\left(\frac{n_\Gamma}{n}\right)$. The terms $n_\Gamma$ and $n_L$ denote the electron concentrations in the direct and indirect valleys, respectively. The recombination coefficients are $R_L = 5.1 \times 10^{-15} cm^3/s$ and $R_\Gamma = 1.3 \times 10^{-10} cm^3/s$ [23].*

$$U_{non-radiative} = C_{nnp} n(np - n_i^2) + C_{ppn} p(np - n_i^2) + \min(n,p)/\tau_{SRH}$$
$$= C_{nnp} n(np - n_i^2) + C_{ppn} p(np - n_i^2) + p/\tau_{SRH}$$

*Equation 2. Non-radiative recombination rate ($U_{non-radiative}$) in terms of the electron density (n), hole density (p) and defect-limited carrier lifetime ($\tau_{SRH}$). Note that since we always presume either undoped or n-type doped material the minority carrier density will always be the hole carrier density (p). The recombination coefficients are $C_{nnp} = 3.0 \times 10^{-32} cm^6/s$ and $C_{ppn} = 7.0 \times 10^{-32} cm^6/s$ [23].*

$$IQE = \frac{U_{radiative}}{U_{radiative} + U_{non-radiative}}$$

*Equation 3. Internal quantum efficiency (IQE) in terms of the radiative recombination rate ($U_{radiative}$) and the non-radiative recombination rate ($U_{non-radiative}$).*

We can consider how the material quality affects the performance of double-heterostructure Ge LEDs, since the defect-limited carrier lifetime ($\tau_{SRH}$) is a strong function of material quality [15], [24], [25]. In our modeling, we assume that the doping level is 1e19 cm$^{-3}$. As shown in Fig. 2, there is no conceivable level of strain that will result in an efficient Ge LED if $\tau_{SRH}$ is less than 1 ns. On the other hand, if $\tau_{SRH}$ can be greater than 10 ns, the IQE can be greater than 50% for large strain values. Given that epitaxially-grown Ge films tend to have $\tau_{SRH}$ of approximately 1 ns [14], [15], considerably less than the bulk lifetime values of >100 ns for the similar level of n-type doping [26], there is an acute need for research efforts such as those of Refs [14]–[16], [27] which explore innovative ways of improving the material quality and thereby improving $\tau_{SRH}$. Without such efforts, an efficient CMOS-compatible LED is not possible no matter how much research efforts would be put into strain engineering and other techniques.

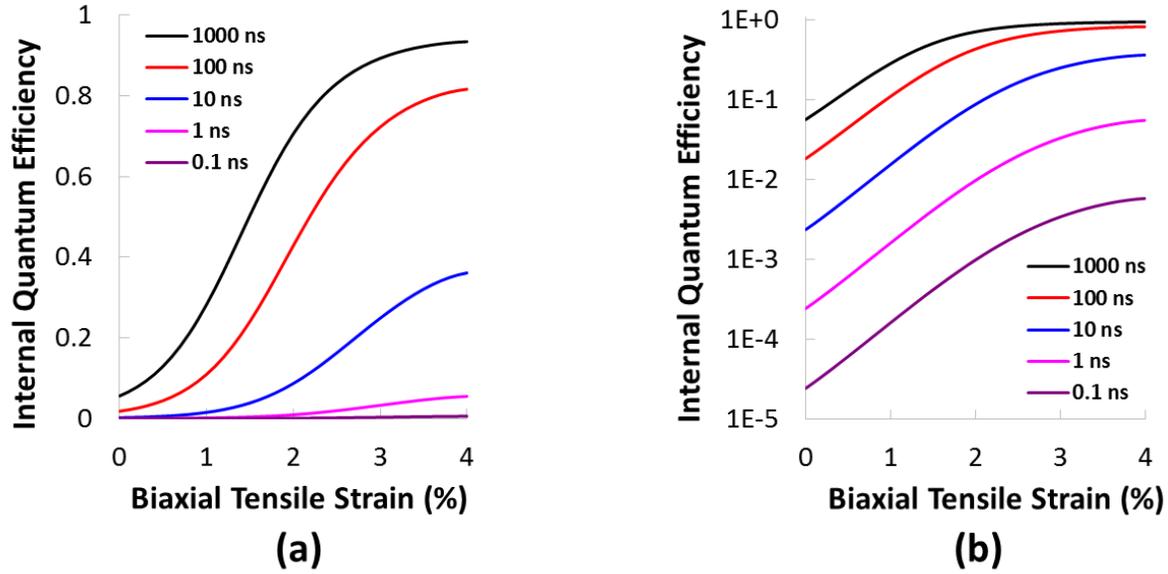

*Fig. 2.* Internal quantum efficiency of a Ge double-heterostructure LED for various $\tau_{SRH}$ values. Doping is assumed to be $1\times 10^{19}$ cm$^{-3}$. (a) Linear scale in y-axis. (b) Logarithmic scale in y-axis.

## *Laser Modeling*

The penultimate goal of using band engineered Ge is to build a low-threshold Ge-on-silicon laser for use in on-chip optical interconnects [3]. Prior modeling has answered some critical questions such as the relative merits of tensile strain and n-type doping [10], the impact of quantum wells [28] and the ideal crystal orientation [29], however the importance of material quality for a low-threshold Ge laser has hitherto been mostly overlooked. To finally investigate the importance of material quality, we extend upon our previous theoretical work [10] by singling out $\tau_{SRH}$ for careful investigation. Our laser modeling in this work employs the exact same approach explained in detail in our previous work [10] with one exception: we have used a corrected absorption coefficient [30] which better accounts for the splitting of the valence bands under biaxial tensile strain.

In the previous section, we showed that maximizing $\tau_{SRH}$, presumably through improved material quality [15], is critically important to the performance of a Ge LED. Unsurprisingly, we find that maximizing the $\tau_{SRH}$ is also very important to achieving a low-threshold Ge laser. As shown in Fig. 3, there is no combination of strain and doping that gives a useable threshold for the case of a defect-limited carrier lifetime below 1 ns. On the other hand, improving $\tau_{SRH}$ to 100 ns makes it possible to achieve thresholds as low as 100 A/cm³. These benefits appear to start saturating once $\tau_{SRH}$ reaches about 100 ns. Given that most epitaxial Ge today has a lifetime of only about 1ns, this means that it is absolutely imperative to improve the material quality if an efficient Ge laser is ever to be realized. Fortunately, this should be within the realm of possibility given that the bulk lifetime of Ge exceeds 100 ns [26].

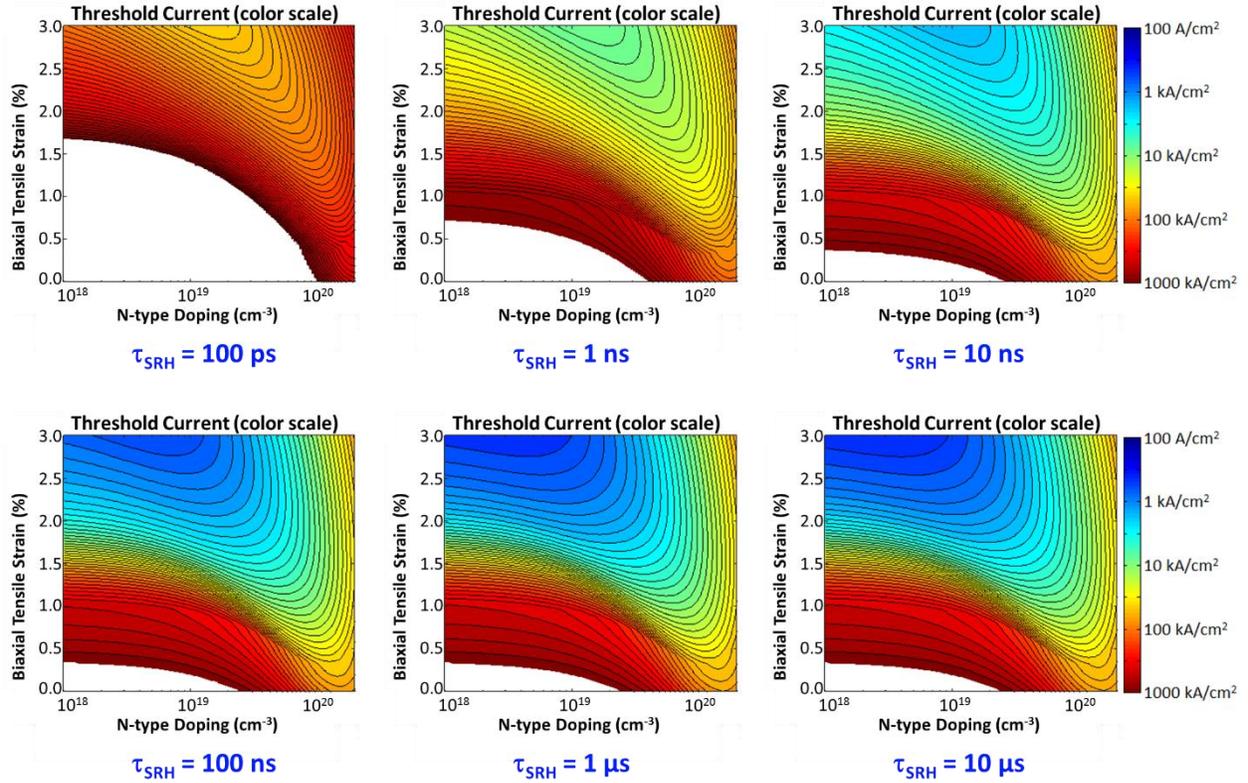

*Fig. 3.* *Threshold current density vs. biaxial strain and n-type doping, shown for various $\tau_{SRH}$ values. In all cases a double heterostructure design with a 300nm thick Ge active region and zero optical cavity loss is assumed.*

Another interesting observation from our laser modeling is that improving $\tau_{SRH}$ appears to offer the most benefit for strain & doping combinations that give the smallest threshold. This is shown explicitly in Fig. 4(a) by plotting the threshold versus strain for several different $\tau_{SRH}$ values, assuming $5\times10^{18}$ cm$^{-3}$ n-type doping. For strains below 1.0%, there is almost no difference between the curves for $\tau_{SRH}$ of 10 ns or greater. At 2.0% strain, the 10-ns curve shows a noticeably higher threshold than the 100 ns – 10 µs curves, and at 3.0% strain even the 100 ns curve has started to diverge from the 1 µs – 10 µs curves. The explanation for this phenomenon is that improving $\tau_{SRH}$ is only helpful when the net carrier lifetime is limited by the defect-assisted recombination process as opposed to other mechanisms such as Auger recombination. In the high threshold regime, the Auger-limited lifetime is about 1 ns or less, and so it makes little difference whether $\tau_{SRH}$ is 1 µs or 10 ns. In the low-threshold regime, however, the Auger-limited lifetime can be closer to 300 ns. In this case, reducing $\tau_{SRH}$ from 1 µs to 10 ns will dramatically reduce the net carrier lifetime and thus increase the threshold by more than an order of magnitude. The key takeaway of this result is that while material quality is already important to improving the performance of Ge lasers, it will become even more important as techniques such as band engineering and n-type doping start to lower lasing thresholds.

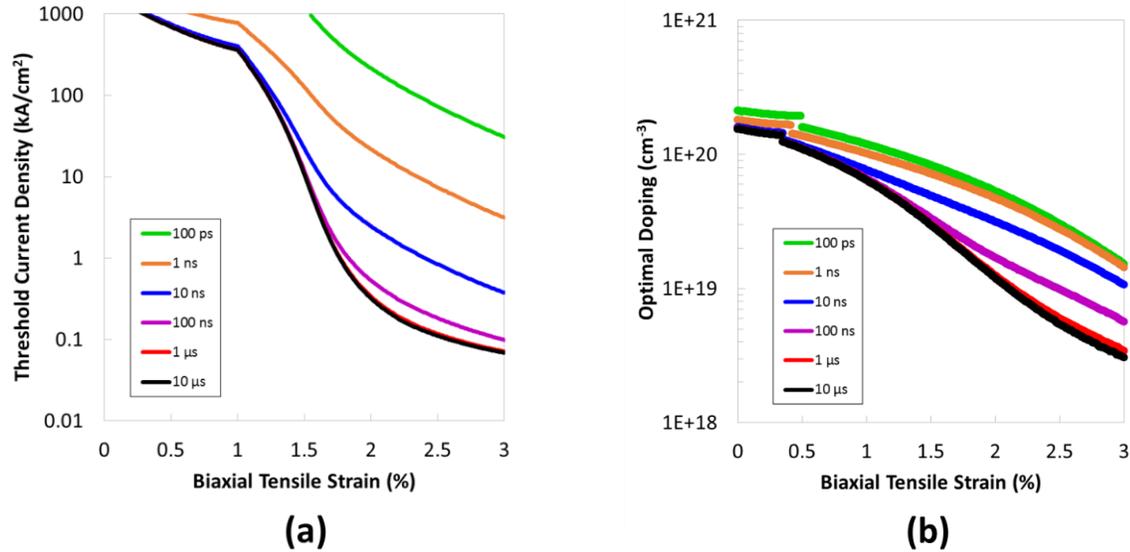

*Fig. 4.* *(a) Threshold current density vs. biaxial strain assuming a 300nm thick Ge active region with $5 \times 10^{18}$ cm$^{-3}$ n-type doping, shown for various $\tau_{SRH}$ values. (b) Optimal doping vs. biaxial strain, shown for various $\tau_{SRH}$ values. In all cases a double heterostructure design with zero optical cavity loss is assumed.*

Another important effect of the defect-limited carrier lifetime is that it also affects the optimal doping level: it is known that too much n-type doping can be harmful [10]. From Fig. 3, we indeed observe that increasing the doping too much will cause the threshold to increase. Also, as can be inferred from plots for various $\tau_{SRH}$ values in Fig. 3, the optimal amount of n-type doping depends quite strongly on the $\tau_{SRH}$. For instance, looking at the top of each plot in Fig. 3, we see that for the case of 3.0% biaxial tensile strain, the threshold is minimized at a doping of more than $1 \times 10^{19}$ cm$^{-3}$ when $\tau_{SRH}$ is 100 ps. But for the case of a 1 µs lifetime, we observe that the threshold is minimized for an n-type doping of less than $5 \times 10^{18}$ cm$^{-3}$. This "optimal doping," i.e. the doping value which minimizes the threshold, is shown explicitly in Fig. 4(b), wherein we consistently observe that the optimal doping value is lower when $\tau_{SRH}$ is higher. On a practical level, what this result means is that n-type doping will become less useful as researchers develop ways to grow Ge with fewer defects and thus longer $\tau_{SRH}$. As shown in Fig. 4(b), assuming 1.5% biaxial strain which has recently been realized in Ge-on-silicon [31], we find for $\tau_{SRH}$ of 1 ns that n-type doping will remain useful for concentrations up to 7.2 x $10^{19}$ cm$^{-3}$. However, if improvements in Ge epitaxy can extend $\tau_{SRH}$ to 100 ns, n-type doping will only be useful at concentrations up to 3.4 x $10^{19}$ cm$^{-3}$ (assuming again 1.5% strain). Such doping levels are significantly lower than what has already been achieved experimentally [32], [33]. Thus, while very heavy n-type doping may improve a Ge laser's threshold now that $\tau_{SRH}$ are quite small [14], as the material quality of Ge improves, it will become necessary to actually reduce the n-type doping from present levels in order to achieve the best possible laser performance.

## *Summary*

In this paper, we show that it is critically important to improve the defect-limited carrier lifetime in order to achieve a high efficiency LED, indicating that much work is needed on improving material quality [15]. Given that state-of-the-art epitaxial Ge has a defect-limited carrier lifetime of only ~1 ns, our modeling suggests that improving this lifetime to ~100 ns would proportionally improve the efficiency of a Ge LED by two orders of magnitude. This makes lifetime improvements even more critical to LED device performance than band engineering such as biaxial strain. Our modeling also shows that improving the defect-limited carrier lifetime is critical for achieving a low-threshold Ge laser. We also show that the defect-limited carrier lifetime will become even more important as techniques such as band engineering start to lower lasing thresholds. This indicates a positive interaction between material quality and tensile strain: increases in the applied tensile strain will make improvements to the carrier lifetime more impactful and vice-versa. Conversely, we observe a negative interaction between material quality and n-type doping: as the material quality improves and reduces the lasing threshold, this will make n-type doping less helpful. This work suggests that to realize a high-efficient Ge LED and a low-threshold Ge laser, we should put more research effort into improving material's quality which has been overlooked so far.

## *Acknowledgements*

This work was supported by the Office of Naval Research (grant N00421-03-9-0002) through APIC Corporation (Dr. Raj Dutt) and by a Stanford Graduate Fellowship. This work was also supported by an INHA UNIVERSITY Research Grant and by the Pioneer Research Center Program through the National Research Foundation of Korea funded by the Ministry of Science, ICT & Future Planning (2014M3C1A3052580). The authors thank Shashank Gupta of Stanford University and Boris M. Vulovic of APIC Corporation for helpful discussions. The authors also thank Ze Yuan of Stanford University for his help implementing the tight-binding code.